 \documentclass[final,5p,times,twocolumn]{elsarticle}

\usepackage{amssymb}
\usepackage{natbib}
\usepackage[utf8]{inputenc}
\usepackage[colorlinks=true,citecolor=blue]{hyperref}

 \usepackage{xcolor}

\journal{Physica E [\href{http://dx.doi.org/10.1016/j.physe.2015.09.004}{Physica E {\bf 75}, 86 (2016)}]}

\newcommand{\text}[1]{\mathsf{#1}}
\newcommand{\kBT}{k_{\rm B}T}
\newcommand{\kB}{k_{\rm B}}
\newcommand{\bea}{\begin{eqnarray}}
\newcommand{\eea}{\end{eqnarray}}
\newcommand{\be}{\begin{equation}}
\newcommand{\ee}{\end{equation}}

\begin{document}

\begin{frontmatter}



\title{Effect of incoherent scattering on three-terminal quantum Hall thermoelectrics
}



\author[icmm]{Rafael S\'anchez}
\address[icmm]{Instituto de Ciencia de Materiales de Madrid (ICMM-CSIC), Cantoblanco 28049, Madrid, Spain}
\author[geneve]{Bj\"orn Sothmann}
\address[geneve]{D\'epartement de Physique Th\'eorique, Universit\'e de Gen\`eve, CH-1211 Gen\`eve 4, Switzerland}
\author[rochester,orange]{Andrew N. Jordan}
\address[rochester]{Department of Physics and Astronomy, University of Rochester, Rochester, New York 14627, U.S.A.}
\address[orange]{Institute for Quantum Studies, Chapman University, Orange, California 92866, U.S.A.}
 
\begin{abstract}
A three-terminal conductor presents peculiar thermoelectric and thermal properties in the quantum Hall regime: it can behave as a symmetric rectifier and as an ideal thermal diode. These properties rely on the coherent propagation along chiral edge channels. We investigate the effect of breaking the coherent propagation by the introduction of a probe terminal. It is shown that chiral effects not only survive the presence of incoherence but they can even improve the thermoelectric performance in the totally incoherent regime.

\end{abstract}


\end{frontmatter}


\section{Introduction}
\label{sec:intro}

The last decades of the 20$^\text{th}$ century saw the development of the field of quantum transport in mesoscopic conductors. The quantum nature of electrical carriers shows up in systems of reduced dimensionality, where their phase coherence is maintained when being transported through the sample. The quantization of conductance in quantum point contacts~\cite{van_wees_quantized_1988,wharam_one-dimensional_1988}, the existence of persistent currents in normal metal rings~\cite{buttiker_josephson_1983,levy_magnetization_1990,jariwala_diamagnetic_2001}, or the possibility to design electronic interferometers~\cite{ji_electronic_2003} are good examples. 
The scattering theory of mesoscopic conductors, developed after the ideas of R. Landauer, has been successfully used in many of these problems, where electron-electron interactions do not play a role. The formal elaboration of the theory was established by M. B\"uttiker by emphasizing the role of multi-terminal measurements~\cite{buttiker_four-terminal_1986}, the magnetic field symmetries~\cite{buttiker_symmetry_1988}, and the importance of decoherence~\cite{buttiker_coherent_1988} and fluctuations~\cite{buttiker_scattering_1992}. For a recent review, see Ref.~\cite{lesovik_scattering_2011}

Another important achievement was the formulation of transport along quantum Hall edge-channels~\cite{halperin_quantized_1982} which is not affected by back-scattering~\cite{buttiker_absence_1988}. The quantum Hall effect manifests in four-terminal measurements in the presence of strong magnetic fields: Together with the longitudinal injection of a current, a transverse resistance is measured that shows plateaus at inverse integer multiples of $h/e^2$~\cite{v._klitzing_new_1980}. 

The effect of interactions can be modeled within the scattering matrix framework by introducing phenomenological probes~\cite{buttiker_coherent_1988}. They consist of one or more terminals whose coupling to the system mimics the desired effect. Voltage~\cite{buttiker_coherent_1988}, dephasing~\cite{de_jong_semiclassical_1996,forster_voltage_2007} or thermometer probes~\cite{engquist_definition_1981,meair_local_2014,bergfield_thermoelectric_2014} can be defined by considering the appropriate boundary conditions on (energy-resolved) charge and heat currents. For example, a voltage probe that injects no net charge current into the system introduces the effect of decoherence by inelastic scattering~\cite{buttiker_coherent_1988}. Electrons are absorbed by the probe and re-injected in the system at a randomized energy. The coupling to the probe defines a crossover between the purely coherent transport and the regime where the electron has lost its phase coherence when propagating between the conductor terminals. It may also lead to enhanced correlations~\cite{texier_effect_2000,oberholzer_positive_2006}. Inelastic scattering is present, e.g., in a conductor coupled to a fluctuating environment. 

The coupling to external fluctuations also generates correlations in the conductor~\cite{goorden_two-particle_2007,goorden_cross-correlation_2008}. They can be a source of transport, even if the conductor itself is in equilibrium. In order to rectify fluctuations from a non-equilibrated environment, the conductor must break electron-hole and left-right symmetries. Such conditions are generally present in mesoscopic circuits and nanojunctions. That is the origin of the mesoscopic Coulomb drag effect~\cite{mortensen_coulomb_2001,levchenko_coulomb_2008,sanchez_mesoscopic_2010}: a current injected in a two terminal conductor generates a current in a second conductor to which it is capacitively coupled~\cite{yamamoto_negative_2006,laroche_positive_2011}. 

A related effect is found in three-terminal conductors if the non-equilibrium situation is induced by a temperature gradient~\cite{sanchez_optimal_2011}. It thus gives rise to a transverse thermoelectric effect. A current is generated between two terminals by the conversion of heat absorbed from the hot third terminal. The hot environment can be fermionic~\cite{sanchez_optimal_2011,sothmann_rectification_2012,sanchez_correlations_2013} or bosonic~\cite{entin-wohlman_three-terminal_2010,ruokola_single-electron_2011,sothmann_magnon-driven_2012,bergenfeldt_hybrid_2014}, provided that it does not inject charge into the system, cf. Ref.~\cite{sothmann_thermoelectric_2015} for a recent review and Refs.~\cite{thierschmann_three-terminal_2015,hartmann_voltage_2015,roche_harvesting_2015} for recent experimental realizations. This effect can be described by a probe terminal which injects heat in the conductor by being maintained at a higher temperature~\cite{jordan_powerful_2013,sothmann_powerful_2013,choi_three-terminal_2015}. The presence of the third probe can also be benefitial for the thermoelectric performance of the conductor~\cite{mazza_thermoelectric_2014}. 

\begin{figure}[t]
\begin{center}
\includegraphics[width=\linewidth,clip]{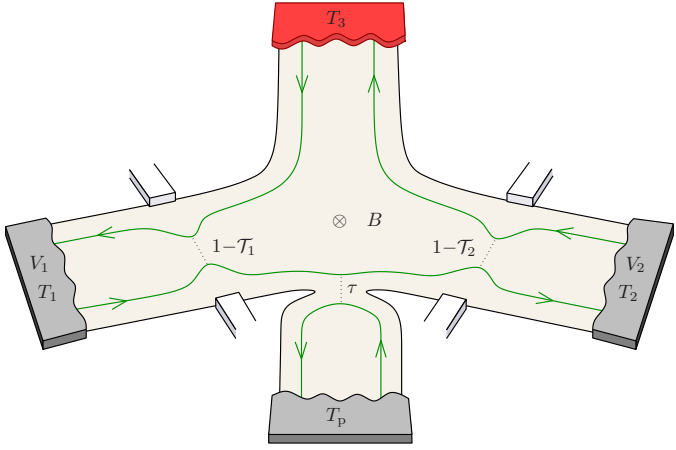}
\end{center}
\caption{\label{scheme} Three-terminal quantum Hall thermoelectric device coupled to a voltage probe. Terminals 1 and 2 hold a charge current in the presence of a voltage bias $V=V_1-V_2$, or a temperature gradient. The latter can either be applied longitudinally (at terminals 1 or 2) or transversally (at terminals 3 or $p$). Terminals 3 and $p$ are considered as probes whose voltage adjusts such that they do not inject charge into the system.  The thermoelectric response relies on the energy dependence of the scattering at the constrictions, in our case quantum point contacts in terminals 1 and 2. The coupling $\tau$ to the probe affects the chiral contributions to the heat conduction.}
\end{figure}

The application of a magnetic field introduces a number of new phenomena~\cite{sothmann_quantum_2014,sanchez_chiral_2015,hofer_quantum_2015,sanchez_heat_2015,bosisio_magnetic_2015, vannucci_interference_2015}. In a recent work, we showed that the quantum Hall effect shows up in the thermoelectric response of a three-terminal configuration~\cite{sanchez_chiral_2015}. The appearance of chiral propagation along edge channels under strong magnetic fields has important consequences in the transverse thermoelectric response of the system~\cite{granger_observation_2009,nam_thermoelectric_2013}. In particular, a finite charge current is predicted in a left-right symmetric conductor as the one represented in Fig.~\ref{scheme}. Furthermore, the system behaves as an ideal thermal diode~\cite{sanchez_heat_2015}. The contributions responsible for these two effects remarkably depend on the coherent propagation between the two conducting terminals.  

In this paper, we address the question of how much these effects are affected by decoherence. We do so by introducing a probe terminal that interrupts the propagation between the two terminals, cf. Fig.~\ref{scheme}. Na\"ively, one expects that the transition to a strongly coupled probe will bring the system to the sequential regime where chiral effects are suppressed. On the other hand, the presence of the probe emphasizes the importance of having a non-equilibrium situation in the middle of the conductor. In our case, it is defined by the left and right moving carriers being thermalized by probes at different temperatures, $T_3$ and $T_{\rm p}$. Hence, we combine the two possible uses of a voltage probe: one of them serves as a model for a non-equilibrium environment able to generate current while the other one acts as a source of decoherence.

\section{Scattering theory}
\label{scattering}

Electronic transport along non-interacting edge channels is well described by the Landauer-B\"uttiker formalism~\cite{buttiker_absence_1988}. We will restrict ourselves to the case with a single edge channel. In this formalism, linear-response charge and heat currents ${\mathbf I}_i=(I_i^e,I_i^h)$ can be expressed in a compact form~\cite{buttiker_four-terminal_1986,sivan_multichannel_1986,butcher_thermal_1990}
\be
\label{Ii}
{\mathbf I}_i=
\frac{1}{h}\sum_j\int dE\left[\delta_{ij}-{\cal T}_{i\leftarrow j}(E)\right]\xi(E)
\left(
	\begin{array}{cc}
		e & eE \\
		E & E^2
	\end{array}
\right)
{\mathbf F}_j,
\ee
in terms of the transmission probabilities ${\cal T}_{i\leftarrow j}(E)$ for electrons injected in terminal $j$ to be absorbed by terminal $i$, and the electric and thermal affinities ${\mathbf F}_j=(F_j^V,F_j^T)$,
with $F_i^V=eV_i/(\kBT)$ and $F_i^T=k_\text{B}\Delta T_i/(\kBT)^2$.
Here $V_i$ and $\Delta T_i$ are the voltage and temperature bias applied to terminal $i=1,2,3,{\rm p}$, respectively, and $\kBT$ is the system temperature. We have introduced the derivative of the Fermi function $\xi(E)=-(\kBT/2)df/dE$. The equilibrium Fermi energy $E_{\rm F}$ is considered in the following as the zero of energy.

Terminals 1 and 2 define the electric conductor which supports a charge current $I^e=I_1^e=-I_2^e$. Terminal 3 models a heat source. Terminal $p$ introduces inelastic scattering. Thus the latter two terminals inject heat but no charge (on average) into the conductor, i.e. $I_3^e=I_{\rm p}^e=0$~\cite{buttiker_coherent_1988}. The voltage of terminals 3 and $p$ are left to accommodate to the configuration at which the probe boundary conditions are satisfied. All other voltages and temperatures are fixed. Charge and heat currents 
\bea
I^e&=GV+\sum_jL_{1j}\Delta T_j\\
I_i^h&=M_{i1}V+\sum_jK_{ij}\Delta T_j
\eea
flow in response to a voltage bias $V=V_1-V_2$ or to a thermal gradient applied to each other terminal. $G$ and $K_{ij}$ are the electrical and thermal conductances. The thermoelectric response is given by the Seebeck and Peltier coefficients, here proportional to $L_{1j}$ and $M_{i1}$, respectively. In the presence of a magnetic field, the linear response coefficients are known to be linked by the Onsager reciprocity relations~\cite{onsager_reciprocal_1931,buttiker_symmetry_1988,butcher_thermal_1990,matthews_experimental_2014}:
\be
L_{1j}(B)=M_{j1}(-B)/T,
\ee
and by energy conservation, $\sum_jI_j^h=0$. 

We will focus here on the transverse Seebeck coefficient, $L_{13}$, and the longitudinal off-diagonal thermal conductances, $K_{12}$ and $K_{21}$. The former term gives rise to an electric current generated between terminals 1 and 2 by conversion of the heat injected from terminal 3 being at a higher temperature. This is the process of relevance for energy harvesting~\cite{sothmann_thermoelectric_2015}. The latter coefficients give information about thermal rectification, i.e. how asymmetrically heat flows along the conductors when the heat source is coupled to terminal 1 or to terminal 2.

The thermoelectric response relies on the presence of energy-dependent scatterers in the conductor which break the electron-hole symmetry. To be specific, we will consider the case with two quantum point contacts that affect the propagation between the conducting terminals and the probes, cf. Fig.~\ref{scheme}. They define constrictions described by a saddle point potential at which electrons can either be transmitted or reflected. Their transmission probability is given by a step function~\cite{fertig_transmission_1987,buttiker_quantized_1990}:
\be
{\cal T}_{l}(E)=\left[1+e^{-2\pi(E-E_l)/\hbar\omega_{l}}\right]^{-1}.
\ee
Both the position of the step, $E_l$, and its broadening, $\hbar\omega_l$, can be tuned by gate voltages. Each junction determines a charge conductance $G_l=[e^2/(h\kBT)]g_l^{(1)}$, 
a thermal conductance $N_l=[h(\kBT)^2]^{-1}g_l^{(3)}$, 
and a thermopower, $S_l=[e/(h\kBT^2G_l)]g_l^{(2)}$, 
all written in terms of the integrals
\be
\label{gnl}
g_{l}^{(n)}=\int dE E^{n-1}{\cal T}_l(E)\xi(E).
\ee
We also define 
\be
\label{jn}
j^{(n)}=\int dE E^{n-1}{\cal T}_1(E){\cal T}_2(E)\xi(E),
\ee
on which the chiral terms depend. It includes the propagation of electrons between the two junctions along the lower edge channel (electrons in the upper channel are absorbed by terminal 3.)
When $E_l-E_{\rm F}\sim\hbar\omega_l,\kBT$, the contact is noisy and leads to an enhanced thermoelectric response~\cite{streda_quantised_1989,molenkamp_quantum_1990,molenkamp_peltier_1992}. Otherwise, it is transparent ($E_l\ll E_{\rm F}$, giving $G_l=e^2/h$), or closed to transport ($E_l\gg E_{\rm F}$). 

The coupling to the probe terminal could also be modeled as a quantum point contact. It allows us to tune its opening and therefore the influence of inelastic scattering. Our interest here is focused on how it breaks the coherent propagation between the two conducting terminals. It will thus be sufficient for that purpose to assume it to be energy independent: ${\cal T}_p(E)=\tau$. General expressions including an arbitrary coupling are given in ~\ref{sec:energydepprobe}.

\section{Incoherent charge transport}
\label{sec:charge}

Breaking the coherent propagation between two barriers in a two terminal conductor leads to a regime where transport is dominated by the sequential scattering at each barrier. In that case, the resistance is given by the series resistance of the two barriers: $G_{\rm seq}^{-1}=G_1^{-1}+G_2^{-1}$. This effect was modeled by B\"uttiker by introducing a voltage probe~\cite{buttiker_coherent_1988}. Note that even in the case where the two junctions are open, the probe can re-emit an electron back to the same reservoir, resulting in a conductance $G_{\rm seq,open}=e^2/(2h)$.

In our setup, terminal 3 acts as a transparent probe for the electrons propagating along the upper branch. Thus, they completely lose coherence in the process. Differently, electrons in the lower branch which are reflected at the probe remain coherent. Thus, the contribution of the terms $j^{(n)}$ will be weighted by a factor $1-\tau$. This is clear in the expression for the charge conductance:
\be
\label{eq:g}
G=G_{\rm seq}\left[(1-\tau)\lambda^{(1)}+\tau\left(1-\frac{G_{\rm seq}}{G_0}\right)\right]^{-1},
\ee
where $\lambda^{(n)}=1-j^{(n)}/(g_{1}^{(n)}+g_{2}^{(n)})$ and the quantum of conductance $G_0=e^2/h$. Tuning the coupling of the probe from $\tau=0$ (closed) to $\tau=1$ (open) we find the crossover from the coherent regime (dominated by $\lambda^{(1)}$~\cite{sanchez_chiral_2015}) to the incoherent regime. 

Note however that we do not recover the sequential result as we have
\be
\label{gt1}
G(\tau{=}1)=\left(G_1^{-1}+G_2^{-1}-G_0^{-1}\right)^{-1}.
\ee
We interpret the additional term given by the quantum of conductance as being due to having two probes. They equilibrate at different voltages because they are coupled to different channels. Electrons that are absorbed by one of the probes are not re-emitted back into the same terminal. Hence, even in the totally incoherent regime, we find a residual chiral effect. Indeed, for an open conductor (${\cal T}_1={\cal T}_2=1$), the extra term $-G_0^{-1}$ in equation~(\ref{gt1}) recovers the quantum Hall conductance $e^2/h$~\cite{buttiker_absence_1988}.

We remark that different kinds of probes give different results. Here we are interested in a voltage probe which injects no current and whose temperature is kept constant. So it in general injects heat. An ideal probe would also inject no heat, and acts as a thermometer, just as a regular probe indicates the local voltage. However, the deviation from the sequential result persists even in this case, as we show in \ref{sec:VTprobe}. 


\section{Thermoelectric response}
\label{sec:thermoel}
\begin{figure}[t]
\begin{center}
\includegraphics[width=0.9\linewidth,clip]{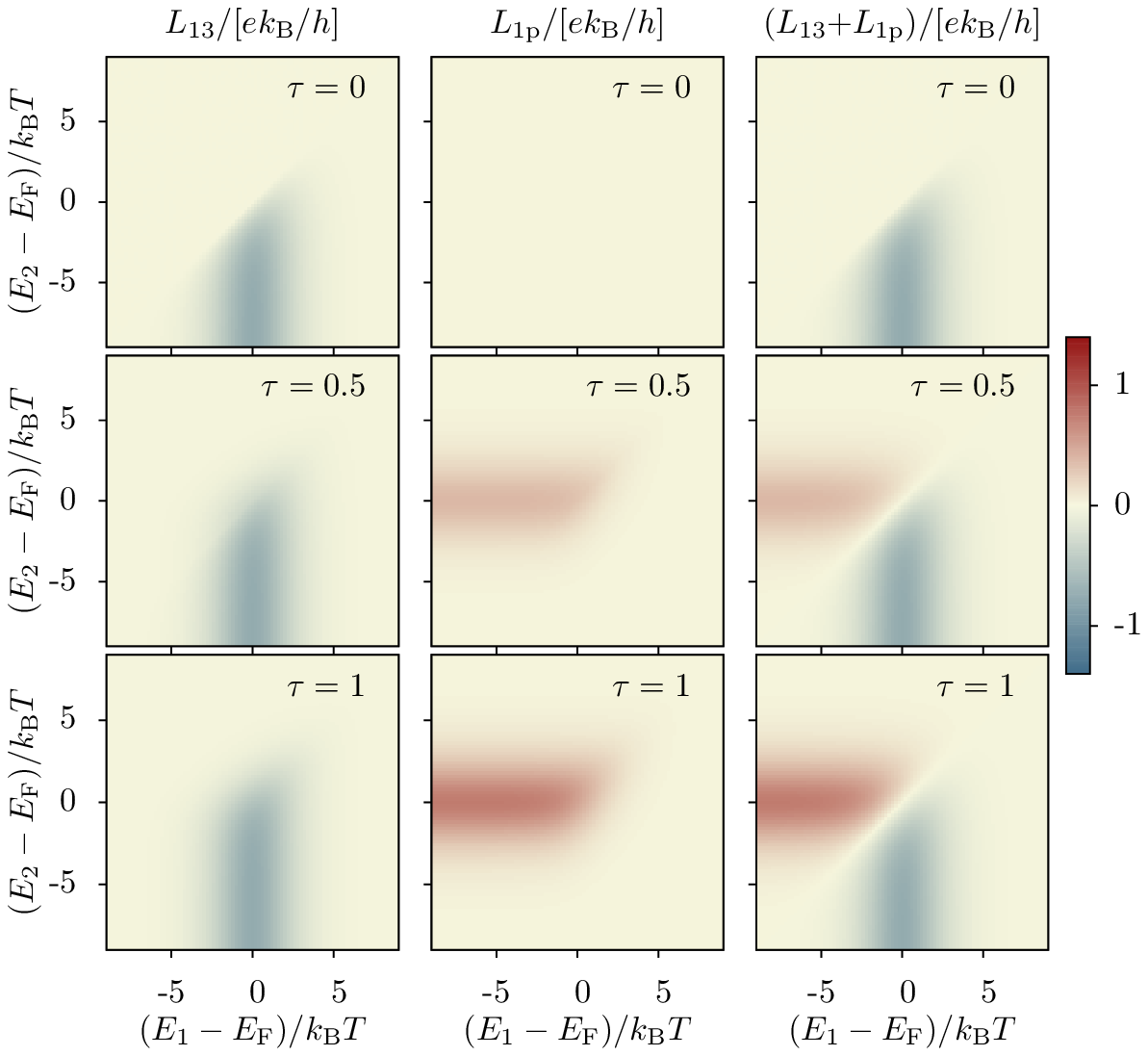}\\
\vspace{0.2cm}
\includegraphics[width=0.9\linewidth,clip]{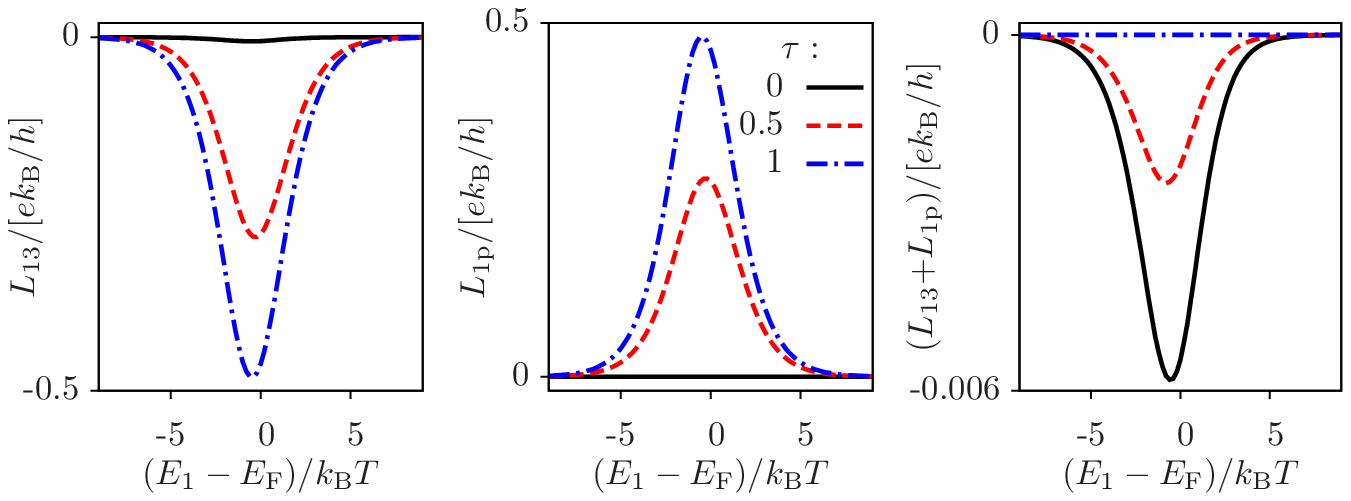}
\vspace{-0.2cm}
\end{center}
\caption{\label{l13p} Transverse thermoelectric responses (Seebeck coefficient) as a function of the threshold energy of the quantum point contacts, $E_l$. The magnetic field is chosen to penetrate the sample as sketched in Fig.~\ref{scheme}. In this case, the temperature gradient is applied to terminal 3 (left column) or the probe (center column). The right column represents to the sum of the two, analogous to the case when the probe is in thermal contact with terminal 3. The different rows correspond to an increasing opening of the probe terminal. The values along the diagonal with symmetric junctions, $E_1=E_2$ are detailed in the lower panels. The QPC step broadening is $\hbar\omega_1=\hbar\omega_2=\kBT/10$.}
\end{figure}

The effect of chirality is more pronounced in the transverse thermoelectric response. If electrons get in contact with the hot terminal while traveling through the conductor, the transverse thermopower (in the absence of a magnetic field) is proportional to the difference of the thermopower of each barrier~\cite{jordan_powerful_2013}, $S_2-S_1$. In the fully coherent case, the presence of edge states introduces a deviation which depends on the terms $j^{(n)}$~\cite{sanchez_chiral_2015}:
\be
{\cal X}_l=\frac{1}{h}\frac{G_lG}{G_1G_2}\left(eTS_{l}j^{(1)}-j^{(2)}\right).
\ee
In consequence, only one of the junctions is necessary in order to have a finite transverse thermoelectric response. This is seen in the upper left panel in Fig.~\ref{l13p}: $L_{13}(\tau{=}0)=0$ if ${\cal T}_1\rightarrow 1$, no matter what the transmission of the right junction is.
  
\begin{figure}[t]
\begin{center}
\includegraphics[width=0.34\linewidth,clip]{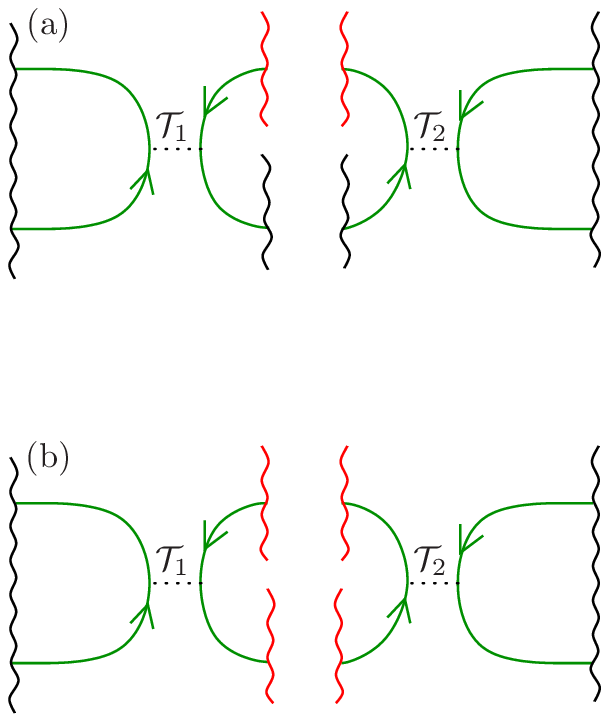}
\quad
\includegraphics[width=0.61\linewidth,clip]{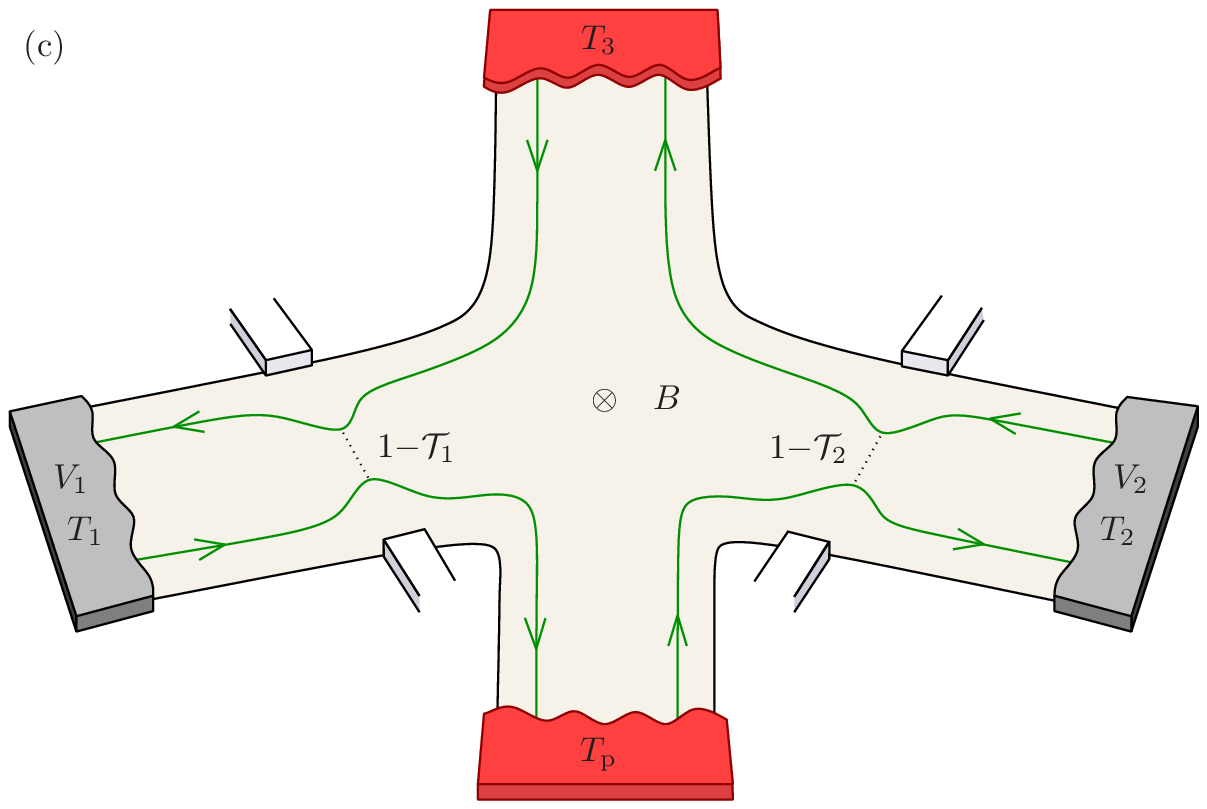}
\end{center}
\caption{\label{schemeSeq} Two different cases are found when the probe is transparent. (a) If the probe is in thermal equilibrium with the conductor, transport is incoherent but chiral. (b) and $($c) If it is in thermal contact with the heat source (in terminal 3), the transverse conductance is analogous to that of the sequential regime, cf. Eq.~(\ref{sseq}).}
\end{figure}

In the presence of the probe, the transverse Seebeck term when terminal 3 is hot reads
\be
\label{eq:l13}
L_{13}(\tau)=G(S_2-S_1)-L_{1{\rm p}}(\tau)+(1-\tau)\frac{e}{\kBT^2}{\cal X}_1,
\ee
where 
\be
L_{1{\rm p}}(\tau)=\tau GS_2
\ee
gives the transverse Seebeck effect when the probe $p$ is hot. 
Note that there is an implicit contribution of the coherent terms in Eq.~(\ref{eq:l13}) through the conductance~(\ref{eq:g}). As expected, the contribution of ${\cal X}_1$ vanishes with the opening of the probe. A totally transparent probe removes the phase coherence of carriers traversing the sample. However, in this limit
\be
\label{eq:l13open}
L_{13}(\tau{=}1)=-GS_1
\ee
only depends on the thermopower of the left junction. The reason is that electrons injected from the hot terminal are either thermalised in terminal 1 or in the probe, cf. Fig.~\ref{schemeSeq}a. They are never absorbed by terminal 2 (which only absorbs electrons thermalised at temperature $\kBT$ in the probe). Even if carriers lose their coherence, the thermoelectric response is still chiral. 

The sequential limit is only recovered if $\tau=1$, and terminal 3 and the probe are in thermal contact, i.e. $\Delta T_3=\Delta T_{\rm p}$. In that case, every electron entering the central region of the system will thermalise at an increased temperature, cf. Fig.~\ref{schemeSeq}b,c. Then, a transverse Seebeck coefficient 
\be
\label{sseq}
S_{\rm seq}=-\frac{L_{13}+L_{1{\rm p}}}{G}=S_1-S_2
\ee
will develop across the conductor. This result is analogous to having a hot cavity in between two cold terminals~\cite{jordan_powerful_2013,sothmann_powerful_2013,choi_three-terminal_2015}. Note however that there is a voltage difference between terminal 3 and the probe. We can interpret equation (\ref{sseq}) in terms of the electron-hole excitations generated in the hot terminals. Those injected from terminal 3 are only partitioned in the left junction. Those injected from the probe are only partitioned at the right junction. Thus, the response is the difference of both contributions. The chiral effect does not manifest because the system is symmetric. Then, our system behaves as a time-reversal conductor, which cannot convert heat when being left-right symmetric, cf. Fig.~\ref{l13p}. Interestingly, in that configuration we obtain an Onsager-like relation $L_{13}(B)+L_{1{\rm p}}(B)=[M_{31}(B)+M_{{\rm p}1}(B)]/T$ without reversing the magnetic field.

The effect of the probe on $L_{13}$ is dramatic in the symmetric configuration with $E_1=E_2$ due to the suppression of the chiral term, cf. Fig.~\ref{l13p}. However, far from this region, the response is practically unaffected. This holds true in particular for the configuration with an open right junction, for which the current generated is the largest.

\subsection{Thermoelectric performance}
Let us discuss the performance of the system as a heat engine when the generated current flows against a load potential. Then, a power $P_l=-VI^e$ is generated with an efficiency $\eta_l=-P_l/I^h_l$, when the temperature gradient is applied to terminal $l$. We calculate the voltage $V_{{\rm m},l}$ that maximizes the power generated. With it, we define the maximal power, $P_{{\rm max},l}=P(V_{{\rm m},l})$ and the efficiency at maximum power: $\eta_{{\rm maxP},l}=\eta(V_{{\rm m},l})$.

\begin{figure}[t]
\begin{center}
\includegraphics[width=\linewidth,clip]{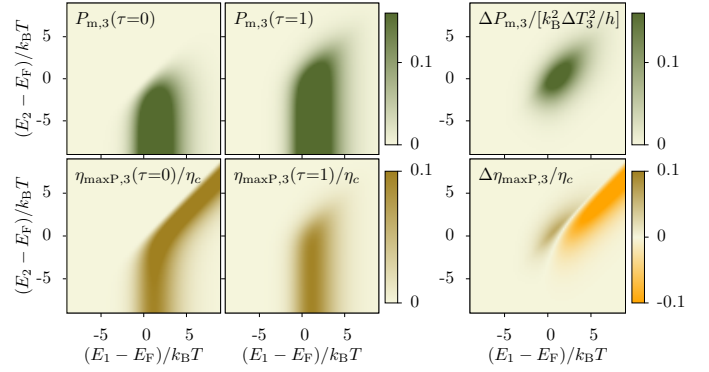}\\
\end{center}
\caption{\label{emp3} Comparison of the maximum power, $P_{{\rm max},3}$, and the efficiency at maximum power, $\eta_{{\rm maxP},3}$ for the cases without and with a transparent probe. The rightmost panels show the difference between them:  $\Delta P_{{\rm max},3}=P_{{\rm max},3}(\tau{=}1)-P_{{\rm max},3}(\tau{=}0)$, and $\Delta{\eta}_{{\rm maxP},3}={\eta}_{{\rm maxP},3}(\tau{=}1)-{\eta}_{{\rm maxP},3}(\tau{=}0)$. }
\end{figure}

As shown in Fig.~\ref{emp3}, the largest $P_{{\rm max},3}$, obtained for $E_2\ll E_{\rm F}$, is the same for the coherent and incoherent cases. In the region where both junctions are noisy, $E_1\approx E_2\approx E_{\rm F}$, we find a competition of two contributions. On one hand, for $\tau=0$ a finite power is due to the term ${\cal X}_1$, only~\cite{sanchez_chiral_2015}. This term vanishes in the presence of incoherence due to the probe. Surprisingly however, as shown in Fig.~\ref{emp3}, the power increases considerably when the probe is transparent. Hence, we find inelastic scattering assisted power generation in a symmetric configuration, which is furthermore of the same order as the chiral and asymmetric one.

Regarding the efficiency, in Ref.~\cite{sanchez_chiral_2015} it was shown that the efficiency at maximum power for $\tau=0$ is maximum close to the symmetric condition $E_1\gtrsim E_2>E_{\rm F}$, cf. Fig.~\ref{emp3}. In the presence of the probe, the efficiency vanishes in that region, cf. Fig.~\ref{emp3}. Interestingly, it increases around the region where the two junctions are half transmitting, coinciding with the inelastic scattering assisted power increase. Remarkably also, in this region $P_{{\rm max},3}$ is close to its maximal value. 

\section{Thermal rectification}
\label{sec:rectif}

Let us discuss the heat flow along the conductor. The possibility to manipulate heat is characterised by the rectification coefficient ${\cal R}_{ij}=K_{ij}/K_{ji}$. We recall that $K_{ij}$ is the heat conductance measured in terminal $i$ when a temperature gradient is applied to terminal $j$. Strong deviations from ${\cal R}_{ij}=1$ indicates that the system behaves as a heat diode. In the linear regime, this is only possible in multiterminal conductors in the presence of a magnetic field~\cite{sanchez_heat_2015,bosisio_magnetic_2015,jiang_phonon-thermoelectric_2015}. 

Let us focus on the case where the two junctions are open, ${\cal T}_1={\cal T}_2=1$. In the fully coherent regime ($\tau=0$), our setup works as an ideal diode, with ${\cal R}_{12}\rightarrow0$~\cite{sanchez_heat_2015}. This is because, if terminal 1 is hot, heat flows without resistance into terminal 2, whereas in the opposite configuration, the heat current is absorbed by terminal 3: $K_{12}=0$.

\begin{figure}[t]
\begin{center}
\includegraphics[width=0.8\linewidth,clip]{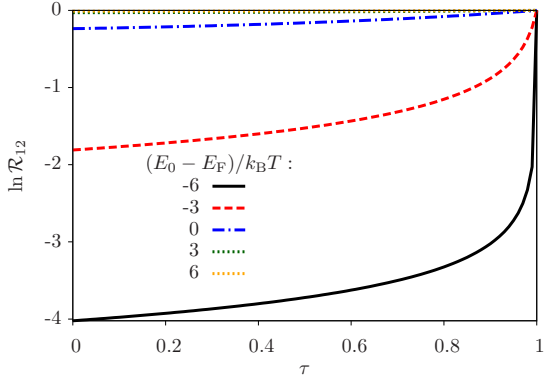}\\
\end{center}
\caption{\label{r12} Rectification coefficient ${\cal R}_{12}$ as a function of the coupling to the probe for different threshold energies of the junctions $E_0=E_1=E_2$. When the probe is transparent ($\tau=1$), no rectification occurs. The heat diode effect appears as the junctions open (for $E_0<E_{\rm F}$).}
\end{figure}

This effect is clearly modified by the probe, which acts as a sink for heat injected in terminal 1. For finite coupling to the probe (and energy independent junctions), we get $K_{21}=-(1-\tau)^2\pi^2(\kBT)^3/(3h)$. If the probe is fully transparent, no electron can propagate between the conducting terminals without being absorbed by a probe. Hence we recover $K_{21}=K_{12}$, i.e. ${\cal R}_{12}=1$. We test the effect of the probe as the two quantum point contacts are simultaneously tuned in Fig.~\ref{r12}. As it is shown there, the diode behaviour is robust against the presence of decoherence: The rectification coefficient is several orders of magnitude smaller than 1, up to transparencies of 99$\%$.


\section{Conclusions}
\label{sec:conclussions}

We have investigated the effect of inelastic scattering on three-terminal quantum Hall thermoelectric conductors. This effect is introduced by means of a voltage probe that interrupts the chiral propagation between the two terminals that define the electrical conductor. The coupling to the probe can be tuned, introducing a phenomenological mechanism for decoherence in the system. We find analytical expressions for the crossover between the coherent and the incoherent regime where transport is sequential. 

Even in the case where electrons totally lose their phase coherence when traversing the system, the response is intrinsically chiral. A signature of this is a finite transverse thermoelectric response in symmetric configurations. The effect of the probe turns out to greatly improve the thermoelectric efficiency. On the other hand, the ideal thermal diode behaviour found in fully coherent configurations is robust in the presence of the probe.

\section*{Acknowledgements}
All the three of us have been postdoc collaborators of Markus B\"uttiker in Geneva. It was there that we started working on three-terminal thermoelectrics in different configurations. It was, however, only soon after his death that we addressed chiral thermoelectrics with edge states via scattering matrix theory and including voltage probes, two of Markus's favorites. Most of what we understand about this, we learnt from him. Many times during this project we have missed his advice and deep insights.
 
We acknowledge financial support from the Spanish MICINN Juan de la Cierva program and MAT2014-58241-P, the COST Action MP1209 and the Swiss National Science Foundation. 

\appendix

\section{Energy dependent coupling to the probe}
\label{sec:energydepprobe}

In this appendix, we present the expressions for the conductance and transverse thermopower in the most general case when the coupling to the probe is energy dependent. Then, it introduces a transmission probability ${\cal T}_{\rm p}(E)$. We use the definitions of Eqs.~(\ref{gnl}) and (\ref{jn}), and define
\be
x_{l}^{(n)}=\int dE E^{n-1}{\cal T}_p(E){\cal T}_l(E)\xi(E)
\ee
and
\be
\tilde j_{l}^{(n)}=\int dE E^{n-1}[1-{\cal T}_p(E)]{\cal T}_1(E){\cal T}_2(E)\xi(E).
\ee
The latter emphasizes that the coherent propagation between junctions 1 and 2 is only possible upon reflection at the probe contact.
With these, we get an electrical conductance:
\be
G=\frac{e^2}{h\kBT}\frac{g_1^{(1)}g_2^{(1)}g_{\rm p}^{(1)}}{\tilde m^{(1)}g_{\rm p}^{(1)}-x_1^{(1)}x_2^{(1)}},
\ee
where $\tilde m^{(1)}=g^{(1)}_1+g^{(1)}_2-\tilde j^{(1)}$. This expression can be used to write simple forms of the thermoelectric coefficients
\be
L_{1{\rm p}}=\frac{\kBT}{e}G\frac{g_{\rm p}^{(1)}x_2^{(2)}-x_2^{(1)}g_{\rm p}^{(2)}}{g_2^{(1)}g_{\rm p}^{(1)}}
\ee
and
\be
L_{13}=G(S_2-S_1)-L_{1{\rm p}}+\frac{e}{\kBT^2}\tilde{\cal X}_1,
\ee
with
\bea
\tilde{\cal X}_1=\frac{\kBT}{e^2}G\left[\frac{g_1^{(2)}\tilde j^{(1)}-g_1^{(1)}\tilde j^{(2)}}{g_1^{(1)}\tilde g_2^{(1)}}\right.\nonumber\\
\left.+\frac{x_2^{(1)}(x_1^{(1)}g_1^{(2)}-g_1^{(1)}x_1^{(2)})}{g_1^{(1)}\tilde g_2^{(1)}g_{\rm p}^{(1)}}\right].
\eea

The case with an energy independent coupling to the probe, considered in the main text, is obtained by replacing $g^{(1)}_{\rm p}=\tau\kBT$, $g^{(2)}_{\rm p}=0$, $x^{(n)}_l=\tau g^{(n)}_l$, and $\tilde j_{l}^{(n)}=(1-\tau)j_{l}^{(n)}$.

\section{Voltimeter and thermometer probe}
\label{sec:VTprobe}

We consider here the case where the probe terminal $p$, on top of not injecting charge, it also does not inject heat into the system, i.e. $I^e_p=I^h_{\rm p}=0$. Its voltage and temperature will accommodate in order to fulfill such boundary conditions. In that case, it can act both as a voltage probe and as a thermometer. We will again restrict to the case where the probe is transparent, ${\cal T}_{\rm p}(E)=1$.

The conductance is then:
\be
\tilde G(\tau{=}1)=\left[G_1^{-1}+G_2^{-1}-(G_0q_\text{H})^{-1}\left(q_\text{H}-G_0TS_1S_2\right)\right]^{-1}.
\ee
In this case, the deviation includes two terms: one depends on the thermopower of the two junctions and the other one only on the quantum of charge and heat conductances, $G_0$ and $q_\text{H}=\pi^2\kB^2T/(3h)$~\cite{pendry_quantum_1983}, respectively. 
If the junctions are energy independent (no thermoelectric effect), the probe will stay at the system temperature. Then the two kinds of probes recover the same result in Eq.~(\ref{gt1}).

For the transverse thermoelectric coefficient, we get:
\be
\tilde L_{13}=(S_2-S_1)\tilde G+\frac{1}{q_\text{H}}S_2(G_1S_1^2-\kB N_1)\tilde G.
\ee
In this case, we recover Eq.~(\ref{eq:l13open}) when the second junction is energy independent, so $S_2=0$.

\end{document}